\documentstyle[titlepage,12pt]{article}
\begin{document}
\renewcommand{\theequation}{\thesection.\arabic{equation}}
\title{Sinh-Gordon,
Cosh-Gordon and Liouville Equations for Strings and Multi-Strings in
Constant
Curvature
Spacetimes.}
\author{A.L. Larsen\thanks{Theoretical Physics Institute, Department
of Physics, University of Alberta, Edmonton, Canada, T6G 2J1.} and N.
S\'{a}nchez \thanks{
Observatoire de Paris,
DEMIRM. Laboratoire Associ\'{e} au CNRS,
UA 336, Observatoire de Paris et
\'{E}cole Normale Sup\'{e}rieure.
61, Avenue de l'Observatoire, 75014 Paris, France.}}
\maketitle
\begin{abstract}
We find that the fundamental quadratic form of classical string
propagation in
$2+1$ dimensional constant curvature spacetimes solves the
Sinh-Gordon
equation, the Cosh-Gordon equation or the Liouville equation. We show
that in
both de Sitter and anti de Sitter spacetimes (as well as in the $2+1$
black
hole anti de Sitter  spacetime), {\it all} three equations must be
included to
cover the generic string dynamics. The generic properties of the
string
dynamics are directly extracted from the properties of these three
equations and
their associated potentials (irrespective of any solution). These  
results
complete and generalize
earlier
discussions
on this topic (until now, only the Sinh-Gordon sector in de Sitter
spacetime
was known).  We also construct new classes of multi-string solutions,  
in terms
of
elliptic
functions, to all three equations in both de Sitter and anti de
Sitter
spacetimes.
Our results can be straightforwardly generalized to constant
curvature
spacetimes of arbitrary dimension, by replacing the Sinh-Gordon
equation, the
Cosh-Gordon equation and the Liouville equation by higher dimensional
generalizations.
\end{abstract}
\section{Introduction and Results}
\setcounter{equation}{0}
In this paper we discuss the dynamics of a relativistic string in
constant
curvature spacetimes, using a combination of geometrical methods and
physical
insight. The kind of problems we are interested in here and the way
of
reasoning, historically had the origin in investigations of the
motion of
vortices in a superfluid \cite{lun,lun1}. Interestingly enough, the
latter
problem, which is equivalent to a theory of dual strings interacting
in a
particular way through a scalar field \cite{lun,lun1}, reduces to
solving two
coupled non-linear partial differential equations, one of which being
a
generalized Sine-Gordon equation. It was soon realized that exactly
the same
equations appear when considering a two-dimensional sigma model
corresponding
to $O(4)$ \cite{poh,lun2,lun3}.
This model, on the other hand, describes a relativistic string in a
(Euclidean
signature)
constant curvature space.

For the theory of fundamental strings, it is important to consider
formulations
in curved spacetimes also, essentially as descriptions of one string
in the
background created by the others. The string equations of motion in
curved
spacetimes are generally highly non-linear in any gauge, which in
most cases
means that the system is non-integrable. Exceptional cases are, among
others,
strings in maximally symmetric spacetimes \cite{zak,eic}. From the
physical
point of view, de Sitter spacetime plays a particular role in this
family of
spacetimes,
since it describes an inflationary universe.
String theory in de Sitter spacetime is therefore also of interest
from the
point of view of cosmic strings and  cosmology and for the open
question of
string self-sustained inflation \cite{ven,ven1}. Specific problems
concerning
the integrability of the equations
describing the dynamics of classical strings in de Sitter spacetime
were
discussed in \cite{nes,san}. The present work is a completion
and generalization of the results presented in those papers.

In Section 2 we set up the general
formalism for classical strings in de Sitter and anti de
Sitter spacetimes, and we derive the equations fulfilled by the
fundamental
quadratic form for a generic string configuration.
The fundamental quadratic form $\alpha(\tau,\sigma)$ is a measure of
the
invariant string size. We show that it solves the Sinh-Gordon
equation, the
Cosh-Gordon equation or the Liouville equation. We find that in order
to cover
the generic string dynamics, {\it all} three equations must be taken
into
account. Associated potentials ($\pm 2\cosh\alpha,\;\pm
2\sinh\alpha,\;\pm
e^\alpha$) to these equations can be respectively defined ($(+)$-sign
for anti
de Sitter spacetime and
$(-)$-sign for de Sitter spacetime). Generic properties of the string
dynamics
are then directly extracted at the level of the equations of motion
from the
properties of these potentials (irrespective of any solution). The
three
equations correspond to three different sectors of the string
dynamics (until
now only the Sinh-Gordon sector (corresponding to the $\cosh\alpha$
potential)
in de Sitter spacetime was known). The differences between the three
sectors in
each spacetime appear mainly for small strings (strings with proper
size
$<1/(\sqrt{2}H)).$

In de Sitter spacetime,
the
Sinh-Gordon sector characterizes the evolution in which small strings
necessarily collapse into a point, while in the Cosh-Gordon sector,
strings
never collapse but reach a minimal size. In anti de Sitter spacetime,
the
situation is exactly the opposite: the Cosh-Gordon sector
characterizes the
evolution in which strings necessarily collapse into a point, while
in the
Sinh-Gordon sector, strings never collapse but reach a minimal size.  
On the
other hand, the dynamics of large strings is rather similar in the  
three
sectors
in each
spacetime (see Figs. 1a, 1b, for instance).

The dynamics of small strings is rather similar in de Sitter and anti
de Sitter
spacetimes, while for large strings (strings with proper size
$>1/(\sqrt{2}H))$
the dynamics is drastically different in the two spacetimes. In de
Sitter
spacetime, the presence of potentials unbounded from below for
positive
$\alpha,$ in all three sectors, makes string instability  
(indefinetely
growing
strings)  unavoidable (in anti de Sitter spacetime, the positive
potential
barriers for positive $\alpha$ prevents the strings from growing
indefinetely).

In Section 3 we present
new classes of explicit solutions in both de
Sitter and anti de Sitter spacetimes, which cover all the three
sectors. These
solutions exhibit the multi-string property
\cite{mik,com,dev,all2}, namely one single world-sheet describes a
finite or
infinite number of different and independent strings. The presence of
multi-strings is a characteristic feature in spacetimes with a
cosmological
constant (constant curvature or asymptotically constant curvature
spacetimes).

In
Section 4, we show that our results also hold for the $2+1$
dimensional black
hole anti de Sitter spacetime \cite{ban}, and we complete earlier
investigation on the dynamics of circular string configurations in
this
spacetime \cite{all1}.

Finally, in Section 5 we give our conclusions.
\section{General Formalism}
\setcounter{equation}{0}
For simplicity, the following analysis is performed for $2+1$
dimensional
spacetimes. However, it is straightforward to generalize the results
to
arbitrary dimensions, following the lines of \cite{san}.

It is well known that $2+1$ dimensional de Sitter spacetime can be
obtained
from flat $R^{1,3}$ spacetime:
\begin{equation}
ds^2(R^{1,3})=-dt^2+du^2+dx^2+dy^2,
\end{equation}
by restricting to the submanifold:
\begin{equation}
\eta_{\mu\nu}q^\mu q^\nu=1,
\end{equation}
where:
\begin{equation}
\eta_{\mu\nu}={\mbox{diag}}(-1,1,1,1),
\end{equation}
\begin{equation}
q^\mu=H(t,u,x,y),
\end{equation}
and $H$ is the Hubble constant of de Sitter spacetime.

Similarly, we can obtain $2+1$ dimensional anti
de Sitter spacetime
from flat $R^{2,2}$ spacetime:
\begin{equation}
ds^2(R^{2,2})=-dt^2-du^2+dx^2+dy^2,
\end{equation}
by restricting to the submanifold:
\begin{equation}
\eta_{\mu\nu}q^\mu q^\nu=-1,
\end{equation}
where:
\begin{equation}
\eta_{\mu\nu}={\mbox{diag}}(-1,-1,1,1),
\end{equation}
\begin{equation}
q^\mu=H(t,u,x,y),
\end{equation}
and $H$ is the Hubble constant of anti de Sitter spacetime.

We can thus treat de Sitter and anti de Sitter spacetimes
simultaneously by
introducing the following notation. We consider a flat spacetime with
line
element $ds^2_{(\epsilon)}$ where $\epsilon=\pm 1:$
\begin{equation}
ds^2_{(\epsilon)}=-dt^2+\epsilon du^2+dx^2+dy^2,
\end{equation}
and restrict to the submanifold:
\begin{equation}
\eta^{(\epsilon)}_{\mu\nu}q^\mu q^\nu=\epsilon,
\end{equation}
where:
\begin{equation}
\eta^{(\epsilon)}_{\mu\nu}={\mbox{diag}}(-1,\epsilon,1,1),
\end{equation}
and $q^\mu$ is in the form of equations
(2.4), (2.8). That is, $\epsilon=+1$ corresponds
to de Sitter spacetime while $\epsilon=-1$ corresponds
to anti de Sitter spacetime.

Let us now consider a bosonic string embedded in the spacetime (2.9).
In the conformal gauge, where the string world-sheet metric is
diagonal,
the Lagrangian is given by:
\begin{equation}
{\cal L}^{(\epsilon)}=\eta^{(\epsilon)}_{\mu\nu}(
\dot{q}^\mu\dot{q}^\nu-
q'^\mu q'^\nu)-\lambda(\eta^{(\epsilon)}_{\mu\nu}q^\mu
q^\nu-\epsilon),
\end{equation}
where the restriction (2.10) has been taken into account
consistently,
through the Lagrange multiplier $\lambda,$ and dot and prime denote
differentiation
with respect to the world-sheet coordinates tau and sigma,
respectively.
The classical string equations of motion and constraints
take then the form:
\begin{equation}
\ddot{q}^\mu-q''^\mu+\epsilon\eta^{(\epsilon)}_{\rho\sigma}
(\dot{q}^\rho
\dot{q}^\sigma-q'^\rho q'^\sigma) q^\mu=0,
\end{equation}
\begin{equation}
\eta^{(\epsilon)}_{\mu\nu}\dot{q}^\mu
q'^\nu=\eta^{(\epsilon)}_{\mu\nu}
( \dot{q}^\mu\dot{q}^\nu+q'^\mu q'^\nu)=0.
\end{equation}
The induced line element on the string world-sheet is given by:
\begin{equation}
dS^2_{(\epsilon)}=\frac{1}{H^2}\eta^{(\epsilon)}_{\mu\nu}dq^\mu
dq^\nu=
-\frac{1}{2H^2}
\eta^{(\epsilon)}_{\mu\nu}( \dot{q}^\mu\dot{q}^\nu-
q'^\mu q'^\nu)(-d\tau^2+d\sigma^2).
\end{equation}
Since we consider only timelike world-sheets, we can define a real
function
$\alpha^{(\epsilon)}$ by:
\begin{equation}
e^{\alpha^{(\epsilon)}}\equiv
-\eta^{(\epsilon)}_{\mu\nu}( \dot{q}^\mu\dot{q}^\nu-
q'^\mu q'^\nu)=-\eta^{(\epsilon)}_{\mu\nu}q^\mu_+ q^\nu_-,
\end{equation}
and we have introduced world-sheet light-cone coordinates
$\sigma_\pm=
(\tau\pm\sigma)/2,$ that is to say, $q^\mu_\pm=\dot{q}^\mu\pm
q'^\mu\;,$ etc.

The fundamental quadratic form, $\alpha^{(\epsilon)},$
is a measure of the invariant string size $S_{(\epsilon)},$
as follows from equations (2.15)-(2.16):
\begin{equation}
S_{(\epsilon)}=\frac{1}{\sqrt{2}H}e^{\alpha^{(\epsilon)}/2}.
\end{equation}
The
string equations of motion and constraints, equations (2.13) and
(2.14),
can now be
written in the more compact form:
\begin{equation}
q^\mu_{+-}=\epsilon e^{\alpha^{(\epsilon)}} q^\mu,
\end{equation}
\begin{equation}
\eta^{(\epsilon)}_{\mu\nu} q^\mu_\pm q^\nu_\pm=0.
\end{equation}

{}From now on we follow the procedure of Refs.\cite{nes,san} (see
also
[1-5,18-20]).
Let us consider the set of vectors:
\begin{equation}
{\cal U}_{(\epsilon)}=
\{ q^\mu, q^\mu_+,q^\mu_-,l^\mu_{(\epsilon)}\},
\end{equation}
where $l^\mu_{(\epsilon)}$ is defined by:
\begin{equation}
l^\mu_{(\epsilon)}\equiv
e^{-\alpha^{(\epsilon)}} e^\mu_{(\epsilon)\rho\sigma\delta}
q^\rho q^\sigma_+ q^\delta_-,
\end{equation}
and $e^\mu_{(\epsilon)\rho\sigma\delta}$ is the completely
anti-symmetric
four-tensor in the spacetime (2.9). It is easy to verify that the
vectors
in ${\cal U}_{(\epsilon)}$
are linearly independent, although not orthonormal. The vector
$l^\mu_{(\epsilon)}$ is normalized according to:
\begin{equation}
\eta_{\mu\nu}^{(\epsilon)}l^\mu_{(\epsilon)}l^\nu_{(\epsilon)}=1.
\end{equation}

The second derivatives of $q^\mu,$ when expressed in the basis
${\cal U}_{(\epsilon)},$ are
given by:
\begin{equation}
q^\mu_{++}=\alpha^{(\epsilon)}_+ q^\mu_+ +u^{(\epsilon)}
l^\mu_{(\epsilon)},
\end{equation}
\begin{equation}
q^\mu_{--}=\alpha^{(\epsilon)}_- q^\mu_- +v^{(\epsilon)}
l^\mu_{(\epsilon)},
\end{equation}
\begin{equation}
q^\mu_{+-}=e^{\alpha^{(\epsilon)}} q^\mu,
\end{equation}
where the functions $u^{(\epsilon)}$ and $v^{(\epsilon)}$ are
implicitly
defined by:
\begin{equation}
u^{(\epsilon)}\equiv
\eta^{(\epsilon)}_{\mu\nu}q^{\mu}_{++}l^\nu_{(\epsilon)},
\;\;\;\;\;\;\;\;\;\;
v^{(\epsilon)}\equiv
\eta^{(\epsilon)}_{\mu\nu}q^{\mu}_{--}l^\nu_{(\epsilon)}.
\end{equation}
{}From these expressions we compute the quantities
$\eta^{(\epsilon)}_{\mu\nu}q^{\mu}_{++-}l^\nu_{(\epsilon)}$ and
$\eta^{(\epsilon)}_{\mu\nu}q^{\mu}_{+--}l^\nu_{(\epsilon)}$ in two
different
ways (using $(q^\mu_{++})_-$ from (2.23) and $(q^\mu_{+-})_+$ from
(2.25), as
well as $(q^\mu_{+-})_-$ from (2.25) and $(q^\mu_{--})_+$ from
(2.24)), and
it then follows that:
\begin{equation}
u^{(\epsilon)}_-=v^{(\epsilon)}_+=0.
\end{equation}
Then, by differentiating equation (2.16) twize, we get:
\begin{equation}
\alpha^{(\epsilon)}_{+-}-\epsilon
e^{\alpha^{(\epsilon)}}+u^{(\epsilon)}
(\sigma_+) v^{(\epsilon)}(\sigma_-) e^{-\alpha^{(\epsilon)}}=0.
\end{equation}
In the previous discussions \cite{nes,san}, it was implicitly assumed
that the
product $u^{(\epsilon)}
(\sigma_+) v^{(\epsilon)}(\sigma_-)$ is positive definite. In that
case the
conformal transformation on the world-sheet metric (2.15):
\begin{eqnarray}
&\alpha^{(\epsilon)}(\sigma_+,\sigma_-)=\hat{\alpha}^{(\epsilon)}
(\hat{\sigma}_+,\hat{\sigma}_-)+\frac{1}{2}\mbox{log}|u^{(\epsilon)}
(\sigma_+)||v^{(\epsilon)}(\sigma_-)|,&\nonumber\\
&\hat{\sigma}_+=\int\sqrt{|u^{(\epsilon)}
(\sigma_+)}|\;d\sigma_+,\;\;\;\;\;\;\;\;
\hat{\sigma}_-=\int\sqrt{|v^{(\epsilon)}
(\sigma_-)}|\;d\sigma_-,&
\end{eqnarray}
which transforms $dS^2_{(\epsilon)}\rightarrow
d\hat{S}^2_{(\epsilon)}:$
\begin{equation}
dS^2_{(\epsilon)}=\frac{-2}{H^2}e^{\alpha^{(\epsilon)}}d\sigma_+
d\sigma_-\;
\rightarrow\;\frac{-2}{H^2}e^{\hat{\alpha}^{(\epsilon)}}
d\hat{\sigma}_+
d\hat{\sigma}_-=d\hat{S}^2_{(\epsilon)},
\end{equation}
reduces equation (2.28) to the equation:
\begin{equation}
\alpha^{(\epsilon)}_{+-}-\epsilon e^{\alpha^{(\epsilon)}}+
e^{-\alpha^{(\epsilon)}}=0.
\end{equation}
This equation is the Sinh-Gordon equation in the case of de Sitter
spacetime
$(\epsilon=+1),$ and the Cosh-Gordon equation in the case of
anti de Sitter spacetime
$(\epsilon=-1).$

It must be noticed, however, that for a generic string world-sheet,
the product
$u^{(\epsilon)}
(\sigma_+) v^{(\epsilon)}(\sigma_-)$ is neither positive nor negative
definite.
In fact, in the next section we shall construct explicit solutions
to the string equations of motion and constraints (2.18)-(2.19)
corresponding
to $u^{(\epsilon)}
(\sigma_+) v^{(\epsilon)}(\sigma_-)$ positive, $u^{(\epsilon)}
(\sigma_+) v^{(\epsilon)}(\sigma_-)$ negative and
$u^{(\epsilon)}
(\sigma_+) v^{(\epsilon)}(\sigma_-)$ identically zero, in {\it both}
de Sitter and anti de Sitter spacetimes.

In the case that
$u^{(\epsilon)}
(\sigma_+) v^{(\epsilon)}(\sigma_-)$ is negative, the conformal
transformation
(2.29) reduces equation (2.28) to:
\begin{equation}
\alpha^{(\epsilon)}_{+-}-\epsilon e^{\alpha^{(\epsilon)}}-
e^{-\alpha^{(\epsilon)}}=0,
\end{equation}
and including also the case when
$u^{(\epsilon)}
(\sigma_+) v^{(\epsilon)}(\sigma_-)=0,$ we conclude that the most
general
equation fulfilled by the fundamental quadratic form
$\alpha^{(\epsilon)}$
is:
\begin{equation}
\alpha^{(\epsilon)}_{+-}-\epsilon e^{\alpha^{(\epsilon)}}+
Ke^{-\alpha^{(\epsilon)}}=0,
\end{equation}
where:
\begin{equation}
K=\left\{ \begin{array}{l}
+1,\;\;\;\;\;\;u^{(\epsilon)}(\sigma_+) v^{(\epsilon)}(\sigma_-)>0 \\
-1,\;\;\;\;\;\;u^{(\epsilon)}(\sigma_+) v^{(\epsilon)}(\sigma_-)<0 \\
\;0,\;\;\;\;\;\;\;\;u^{(\epsilon)}(\sigma_+)
v^{(\epsilon)}(\sigma_-)=0
\end{array}\right.
\end{equation}
and:
\begin{equation}
\epsilon=\left\{ \begin{array}{l}
+1,\;\;\;\;\;\;{\mbox{de Sitter}} \\  -1,\;\;\;\;\;\;{\mbox{anti de
Sitter}}
\end{array}\right.
\end{equation}
Equation (2.33) is either the Sinh-Gordon equation ($\epsilon=K=\pm
1$),
the Cosh-Gordon equation ($\epsilon=-K=\pm 1$) or the
Liouville equation ($K=0$), and all three equations appear in both de
Sitter
and
anti de Sitter spacetimes. This does not mean, of course, that the
string
dynamics is the same in de Sitter and anti de Sitter spacetimes.

Let us
define a potential $V^{(\epsilon)}(\alpha^{(\epsilon)})$ by:
\begin{equation}
\alpha^{(\epsilon)}_{+-}+\frac{dV^{(\epsilon)}(\alpha^{(\epsilon)})}
{d\alpha^{(\epsilon)}}=0,
\end{equation}
(so that if $\alpha^{(\epsilon)}=\alpha^{(\epsilon)}(\tau),$ then
$\;\frac{1}{2}(\dot{\alpha}^{(\epsilon)})^2+V^{(\epsilon)}
(\alpha^{(\epsilon)})=$ const.). Then, it
follows that in the case of de Sitter spacetime:
\begin{equation}
V^{(+1)}(\alpha)=\left\{ \begin{array}{l}
-2\cosh\alpha,\;\;\;\;\;\;K=+1 \\
-2\sinh\alpha,\;\;\;\;\;\;K=-1\\
\;\;\;\;\;-e^{\alpha},\;\;\;\;\;\;\;\;\;\;\;K=0 \end{array}\right.
\end{equation}
while in the case of anti de Sitter spacetime:
\begin{equation}
V^{(-1)}(\alpha)=\left\{ \begin{array}{l}
2\sinh\alpha,\;\;\;\;\;\;K=+1
\\ 2\cosh\alpha,\;\;\;\;\;\;K=-1\\
\;\;\;\;\;e^{\alpha},\;\;\;\;\;\;\;\;\;\;\;K=0 \end{array}\right.
\end{equation}
and we have skipped the $(\pm)$-index on $\alpha.$ Notice that to the
Sinh-Gordon
equation corresponds the $\cosh\alpha$ potential and vice-versa.

The results (2.36)-(2.38) are represented in Fig.1., showing the
different
potentials in
the
cases of de Sitter and anti de Sitter spacetimes, respectively.
Until now only the $K=+1$ sector in de Sitter spacetime was known.
The new
features introduced by the new sectors $K=0$ (corresponding to the
Liouville
equation) and $K=-1$ (corresponding to the Cosh-Gordon equation in
the case of
de Sitter spacetime and to the Sinh-Gordon equation  in the case of  
anti
de Sitter
spacetime) appear for negative $\alpha$ ("small" strings). Small
strings with
proper size $<1/(\sqrt{2}H)$ in the $K=-1$ sector (inside the horizon
in the
case of de Sitter spacetime), do not collapse into a point (as is the
case in the
$K=+1$ sector) but have a minimal size.

The main differences between
de Sitter
and anti de Sitter potentials are for positive $\alpha$ (strings with
proper
size $>1/(\sqrt{2}H).$
In the case of de Sitter spacetime (Fig.1a.),
the potentials are unbounded from below
for large strings (large positive $\alpha$), while for small strings
(large negative $\alpha$) they are either growing indefinetely, flat
or
unbounded from below. In the case of anti de
Sitter spacetime (Fig.1b.), on the other hand, the potentials grow
indefinetely
for large strings (large positive $\alpha$), while for small strings
(large negative $\alpha$) they are either growing indefinetely, flat
or
unbounded from below.

{}From these results we can deduce the generic features of strings
propagating in de Sitter and anti de Sitter spacetimes: Large strings
(large positive $\alpha$) in de Sitter spacetime generically expand
indefinitely, while small strings (large negative $\alpha$) either
bounce or
collapse. In anti de Sitter spacetime, large strings generically
contract, while small strings either bounce or collapse.
For small strings (large negative $\alpha$) the
dynamics is similar in de Sitter and anti de Sitter spacetimes, while
for large
strings (large positive $\alpha$)
it is completely different in the two spacetimes.

Notice that the $\epsilon$ in equation (2.33), which distinguishes
between
de Sitter and anti de Sitter spacetimes, corresponds to the
$"K"$ in the notation of
Ref.\cite{nes}. Our $K$ in equation (2.33) was missed in
Refs.\cite{nes,san}; only
the solutions corresponding to $K=+1$ were found there.
\section{Explicit Examples}
\setcounter{equation}{0}
The exact ("global", i.e.  the whole world-sheet)
solutions to the string equations of motion and constraints in de
Sitter and anti de Sitter spacetimes considered in the literature
until now [12-15, 17, 21-23], describe different classes of string
solutions of
generic shape, circular strings and stationary strings. These
solutions exhibit
the multi-string property
\cite{mik,com,dev,all2}, namely one single world-sheet describes a
finite or
infinite number of different and independent strings. The presence of
multi-strings is a characteristic feature in spacetimes with a
cosmological
constant (constant curvature or asymptotically constant curvature
spacetimes).
All these solutions fall in the $K=+1$ sector, i.e. are solutions to  
the
Sinh-Gordon equation in
the case of de Sitter spacetime and to the Cosh-Gordon equation in
the case
of anti de Sitter Spacetime. We shall now
construct larger families of exact solutions which fall into
{\it all} three sectors $K=\pm 1,\;0.$

Consider first the following algebraic problem:
What is the most general ansatz which reduces the string equations of
motion
and constraints to {\it ordinary} differential equations, in
spacetimes of
the form:
\begin{equation}
ds^2=-a(r)dt^2+\frac{dr^2}{a(r)}+r^2 d\phi^2.
\end{equation}
The string equations of motion
are given by:
\begin{eqnarray}
\ddot{t}\hspace*{-2mm}&-&\hspace*{-2mm}t''+\frac{a_{,r}}{a}
(\dot{t}\dot{r}-
t'r')=0,\nonumber\\
\ddot{r}\hspace*{-2mm}&-&\hspace*{-2mm}r''-\frac{a_{,r}}{2a}
(\dot{r}^2-r'^2)+
\frac{aa_{,r}}{2}(\dot{t}^2-t'^2)-ar
(\dot{\phi^2}-\phi'^2)=0,\nonumber\\
\ddot{\phi}\hspace*{-2mm}&-&\hspace*{-2mm}\phi''+\frac{2}{r}
(\dot{\phi}
\dot{r}-\phi' r')=0,
\end{eqnarray}
while the constraints take the form:
\begin{eqnarray}
\hspace*{-2mm}&-&\hspace*{-2mm}a(\dot{t}^2+t'^2)+\frac{1}{a}
(\dot{r}^2+r'^2)
+r^2(\dot{\phi}^2+\phi'^2)=0,\nonumber\\
\hspace*{-2mm}&-&\hspace*{-2mm}a\dot{t}t'+\frac{1}{a}\dot{r}r'+
r^2\dot{\phi}\phi'=0.
\end{eqnarray}
Since the Christoffel symbols depend only on $r,$ the desired ansatz
is:
\begin{equation}
r=r(\xi^1),\;\;\;\;t=t(\xi^1)+c_1 \xi^2,\;\;\;\;\phi=\phi(\xi^1)+c_2
\xi^2,
\end{equation}
where $(\xi^1, \xi^2)$ are the two world-sheet coordinates (one of
which is
timelike, the other spacelike), and $(c_1, c_2)$ are two arbitrary
constants.
With this ansatz, equations (3.2) are solved by:
\begin{equation}
\frac{dt}{d\xi^1}=\frac{k_1}{a(r)},\;\;\;\;\;\;\;\;\;\;
\frac{d\phi}{d\xi^1}=\frac{k_2}{r^2},
\end{equation}
\begin{equation}
\left( \frac{dr}{d\xi^1}\right)^2=-a(r)
r^2c_2^2-\frac{a(r)}{r^2}k_2^2+k_1^2+
a^2(r) c_1^2,
\end{equation}
where $(k_1, k_2)$ are two integration constants. For the
constraints,
equations (3.3), to be
fulfilled, we must have:
\begin{equation}
k_1 c_1=k_2 c_2.
\end{equation}
In particular, circular string dynamics as considered in
[12-15, 17, 21, 23] corresponds to
$c_1=k_2=0$ and $(\xi^1,\xi^2)=(\tau,\sigma),$ while the infinitely
long
stationary strings considered in
[15] correspond to the "dual" choice
$c_2=k_1=0$ and $(\xi^1,\xi^2)=(\sigma,\tau).$

The induced line element on the string world-sheet is:
\begin{equation}
dS^2=(r^2 c_2^2-a(r)c_1^2)[-(d\xi^1)^2+(d\xi^2)^2],
\end{equation}
such that the fundamental quadratic form is given by:
\begin{equation}
e^\alpha=2|r^2 c_2^2-a(r)c_1^2|.
\end{equation}

Let us now return to our main interest here: strings in de Sitter and
anti de
Sitter spacetimes. In this case, the function $a(r)$ is given by:
\begin{equation}
a_{(\epsilon)}=1-\epsilon H^2 r^2.
\end{equation}
In the case of anti de Sitter spacetime $(\epsilon=-1),$
the static coordinates $(t,r,\phi)$ cover the complete
manifold, while for de Sitter spacetime $(\epsilon=+1),$ they cover
only the
region inside the horizon; the complete de Sitter manifold can
however be
covered by four coordinate patches of the form (3.1), (3.10), see for
instance
\cite{rin}. Notice that the
equation (3.6) for the radial coordinate can be solved explicitly in
terms of the Weierstrass elliptic $\wp$-function \cite{abr}.
The other two equations (3.5)
can then be integrated; the results being expressed in terms of
the Weierstrass elliptic $\sigma$ and $\zeta$-functions \cite{abr}.
We have thus solved completely the string equations of motion and
constraints
using the ansats (3.4) in both de Sitter and anti de Sitter
spacetimes, but
the explicit
expressions of the solutions are not important here. It should be
also stressed that in general the ansatz (3.4)  does not lead to
solutions
automatically
fulfilling the standard closed or open string boundary conditions,
see for
instance \cite{gsw}. However, imposing the
boundary conditions does not arise any problem. In some cases the
ansats
(3.4) actually {\it does} lead to solutions fulfilling the standard
boundary
conditions; an example is $c_1=k_2=0,$ in which case the solution
describes
dynamical circular strings [12-15, 17, 21, 23]. Finally, we are often
interested
in
string
solutions that do not fulfill the  standard closed or open string
boundary
conditions; this is for instance the case for infinitely
long strings \cite{all2,zel} or finite open
strings with external forces acting on the endpoints
of the strings \cite{vil2,fro}.

Let us consider the spacetime
region where $(c_2^2+\epsilon H^2 c_1^2)r^2\geq c_1^2$
(similar conclusions are reached in the other region). In this case
$\xi^1$ is
the timelike world-sheet coordinate, $\xi^1\equiv\tau/H.$
Then, equations (3.6), (3.9)
lead to:
\begin{eqnarray}
\left( \frac{d\alpha^{(\epsilon)}}{d\tau}\right) ^2-2\epsilon
e^{\alpha^{(\epsilon)}}+\frac{8}{H^2}[c_1^2 c_2^2
\hspace*{-2mm}&-&\hspace*{-2mm}(c_2^2+\epsilon H^2 c_1^2)
(k_1^2+\epsilon H^2 k_2^2)]
e^{-\alpha^{(\epsilon)}}\nonumber\\
=\hspace*{-2mm}&-&\hspace*{-2mm}\frac{4}{H^2}(c_2^2-\epsilon H^2
c_1^2).
\end{eqnarray}
Now, by tuning the
constants of motion to fix the sign
of the square bracket, and by performing conformal transformations
of the form (2.28), we can, after differentiation with respect to  
tau,  reduce
this equation to either
the Sinh-Gordon equation, the Cosh-Gordon equation or the Liouville
equation:
\begin{eqnarray}
&\epsilon[c_1^2 c_2^2-(c_2^2+\epsilon H^2 c_1^2)
(k_1^2+\epsilon H^2
k_2^2)]<0&\;\;\;\;\Rightarrow\;\;\;\;\;\;\mbox{Sinh-Gordon}
\nonumber\\
&\epsilon[c_1^2 c_2^2-(c_2^2+\epsilon H^2 c_1^2)
(k_1^2+\epsilon H^2
k_2^2)]>0&\;\;\;\;\Rightarrow\;\;\;\;\;\;\mbox{Cosh-Gordon}
\nonumber\\
&\;[c_1^2 c_2^2-(c_2^2+\epsilon H^2 c_1^2)
(k_1^2+\epsilon H^2 k_2^2)]=0\;&\;\;\;\;\Rightarrow\;\;\;\;\;\;
\mbox{Liouville}\nonumber
\end{eqnarray}
Thus, we have constructed explicit solutions to the string equations
of
motion and constraints associated to
the Sinh-Gordon equation, the Cosh-Gordon equation or the Liouville
equation
and all three equations appear in both de Sitter and anti de Sitter
spacetimes.

We close this section with the following remark: The ansatz (3.4) is
a
generalization of both the circular string ansatz
($c_1=0,\;\;\phi(\xi^1)=\mbox{const.},\;\;\xi^1$ timelike) and the
stationary
string ansatz ($c_2=0,\;\;t(\xi^1)=\mbox{const.},\;\;\xi^1$
spacelike). In both
these cases, it was shown in Refs. [12-15] that the resulting
solutions in de Sitter and anti de Sitter spacetimes should be
interpreted as
multi-string solutions, that is to say, string solutions where one
single
world-sheet
describes finitely or infinitely many different and independent
strings. The
existence of such
multi-string solutions appears to be a quite general feature in
constant
curvature (and asymptotically constant curvature) spacetimes.
\section{The 2+1 BH-ADS Spacetime}
\setcounter{equation}{0}
As another example to illustrate our general results of Section 2, we
now
consider the $2+1$ dimensional black hole anti de Sitter spacetime
(BH-ADS).

The metric of the $2+1$ dimensional BH-ADS spacetime in its standard
form
is given by \cite{ban}:
\begin{equation}
ds^2=(\frac{J^2}{4r^2}-\Delta)dt^2+\frac{dr^2}{\Delta}-Jdt dr +r^2
d\phi^2,
\end{equation}
where:
\begin{equation}
\Delta=\frac{r^2}{l^2}-M+\frac{J^2}{4r^2}.
\end{equation}
Here $M$ represents the mass, $J$ is the angular momentum and the
cosmological
constant is $\Lambda=-1/l^2.$ For $M^2 l^2\geq J^2,$ there are two
horizons
$(g_{rr}=\infty):$
\begin{equation}
r^2_\pm=\frac{Ml^2}{2}\left( 1\pm\sqrt{1-\frac{J^2}{M^2
l^2}}\;\right),
\end{equation}
and a static limit $(g_{tt}=0):$
\begin{equation}
r^2_{\mbox{st}}=M l^2.
\end{equation}
This spacetime has attracted a lot of interest recently (for a
review, see
for instance \cite{car}), since the causal structure is similar to
that of the
four dimensional Kerr spacetime. However, notice that there is no
strong curvature singularity at $r=0,$ in fact:
\begin{equation}
R_{\mu\nu}=2\Lambda g_{\mu\nu},
\end{equation}
that is to say, the curvature is constant everywhere and the
spacetime
is locally and asymptotically isometric to $2+1$ dimensional
anti de Sitter spacetime; this is of course why it is also relevant
for our
purposes here. For more details on the local and global geometry of
the BH-ADS
spacetime, see for instance Refs. [16, 30-32].

The problem of the string propagation in the BH-ADS spacetime
was completely analyzed and the circular string motion was exactly
solved, in
terms of elliptic functions,
by the present authors in \cite{all1}. The
equation determining the string loop radius as a function of time is:
\begin{equation}
\left( \frac{dr}{d\tau}\right)^2+r^2\left( \frac{ r^2}{l^2}-M\right)
+
\frac{J^2}{4}-E^2=0,
\end{equation}
where $E^2$ is a non-negative integration constant,
while the fundamental quadratic form $\alpha,$ which
determines the invariant size
of the string, is given by:
\begin{equation}
e^\alpha=2 r^2/l^2.
\end{equation}
It is then straightforward to show that the equation (4.6)
becomes:
\begin{equation}
\left( \frac{d\alpha}{d\tau}\right)^2+2e^\alpha-\frac{8}{l^2}
\left( E^2-\frac{J^2}{4}\right) e^{-\alpha}= 4M.
\end{equation}
After performing a conformal transformation of the form (2.28) and
differentiating with respect to tau,
this equation reduces to the (i) Sinh-Gordon equation if $E^2 <
J^2/4,$ (ii) to
the
Cosh-Gordon equation if $E^2 > J^2/4$ and (iii) to the Liouville
equation if
$E^2=J^2/4,$
thus all three equations are present. Notice finally that the three
different types of allowed dynamics as reported in \cite{all1},
essentially
whether the
circular string collapses into $r=0$ (case (ii)) or not (case (i)),
precisely
correspond
to these different equations (in the limiting case (iii), the string
contracts
from the static limit to $r=0$).
\section{Concluding Remarks}
\setcounter{equation}{0}
In conclusion, we have shown that the fundamental quadratic form of
classical
string propagation in $2+1$ dimensional constant curvature spacetimes
solves
the Sinh-Gordon equation, the Cosh-Gordon equation or the Liouville
equation.
We have shown that in  both de Sitter and anti de Sitter spacetimes
(as well as
in the $2+1$ BH-ADS spacetime), all three equations must be included
to cover
the generic string dynamics.  This is particularly enlightening,
since generic
features of the string propagation in these spacetimes can be read
off directly
at the level of the equations of motion from the properties of the
Sinh,
Cosh and Liouville potentials, without need of solving the
equations.
We also constructed new classes of  explicit solutions to
{\it all} three equations in
both de Sitter and anti de Sitter spacetimes, exhibiting the
multi-string
property.

Finally it is worth to observe that our results suggest the existence  
of
various kinds of dualities relating the different string solutions in  
de Sitter
and anti de Sitter spacetimes. From the potentials, Eqs.  
(2.36)-(2.38), it
follows, in particular, that small strings are dual  
($\alpha\rightarrow
-\alpha$) to large strings in the $K=+1$ ($K=-1$) sector of de Sitter  
(anti de Sitter) 

spacetime. Furthermore, small (large) strings in the $K=-1$ sector in  
de Sitter
spacetime are dual ($\alpha\rightarrow  
-\alpha,\;\;\epsilon\rightarrow
-\epsilon$) to large (small) strings in the $K=+1$ sector in anti de  
Sitter
spacetime.
\vskip 12pt
\hspace*{-6mm}{\bf Acknowledgements:}\\
The work by A.L. Larsen was supported by NSERC (National
Sciences and Engineering Research Council of Canada). We also
acknowledge
support from NATO
Collaboration Grant CRG 941287.

\newpage
\begin{centerline}
{\bf Figure Caption}
\end{centerline}
\vskip 24pt
\hspace*{-6mm}Fig.1. The potentials (2.37), (2.38) determining the
dynamics of strings in (a) de Sitter spacetime and (b) anti de Sitter
spacetime, respectively.  For each spacetime, the differences between
the three
sectors $(K=\pm 1,\;0$) appear for negative $\alpha$ (i.e. for
strings with
proper size $<1/(\sqrt{2}H)$). The differences between de Sitter and
anti de
Sitter potentials are for positive $\alpha$ (i.e. for strings with
proper size
$>1/(\sqrt{2}H)$). For small strings (large negative $\alpha$) the
dynamics is similar in both de Sitter and anti de Sitter spacetimes,
while for
large strings (large positive $\alpha$)
it is completely different in the two spacetimes.

\newpage
\begin{thebibliography}{11}
\bibitem{lun}M. Rasetti, T. Regge, Physica 80A, (1975) 217.
\bibitem{lun1}F. Lund and T. Regge, Phys. Rev. D14, (1976) 1524.
\bibitem{poh}K. Pohlmeyer, Commun. Math. Phys. 46, (1976) 207.
\bibitem{lun2}F. Lund, Phys. Rev. D15, (1977) 1540.
\bibitem{lun3}F. Lund, Phys. Rev. Lett. 38, (1977) 1175.
\bibitem{zak}V.E. Zakharov and A.V. Mikhailov, JETP 47, (1979) 1017.
\bibitem{eic}H. Eichenherr, {\it in}: Integrable quantum field
theories,
Lecture
             Notes in Physics, vol. 151, Tv\"{a}rminne Proc., ed.
             J. Hietarinta and C. Montonen (Springer-Verlag, Berlin,
1982)
\bibitem{ven}N. S\'anchez and G. Veneziano, Nucl. Phys.
             B333, (1990) 253.
\bibitem{ven1}M. Gasperini, N. S\'anchez and G. Veneziano,\\
             IJMP A6, (1991) 3853; Nucl. Phys.  B364, (1991) 365.
\bibitem{nes}B.M. Barbashov and V.V. Nesterenko, Commun. Math. Phys.
78, (1981)
             499.
\bibitem{san}H.J. de Vega and N. S\'{a}nchez, Phys. Rev. D47, (1993)
3394.

\bibitem{mik}H.J. de Vega, A.V. Mikhailov and N. S\'{a}nchez, Mod.
Phys. Lett.
             A, Vol 9, No 29 (1994) 2745; Teor. Mat. Fiz. 94, (1993)
232.
\bibitem{com}F. Combes, H.J. de Vega, A.V. Mikhailov and N.
S\'{a}nchez,
             Phys. Rev. D50, (1994) 2754.
\bibitem{dev}H.J. de Vega, A.L. Larsen and N. S\'{a}nchez, Nucl.
Phys. B427,
             (1994) 643.
\bibitem{all2}A.L. Larsen and N. S\'{a}nchez, Phys. Rev. D51, (1995)
6929.
\bibitem{ban}M. Banados, C. Teitelboim and J. Zanelli, Phys. Rev.
Lett. 69,
             (1992) 1849.
\bibitem{all1}A.L. Larsen and N. S\'{a}nchez, Phys. Rev. D50, (1994)
7493.
\bibitem{omn}R. Omnes, Nucl. Phys. B149, (1979) 269.
\bibitem{nes2}B.M. Barbashov, V.V. Nesterenko and A.M. Chervjakov,
              Lett. Math. Phys. 3, (1979) 359.
\bibitem{nes3}B.M. Barbashov, V.V. Nesterenko and A.M. Chervjakov,
              Teor. Mat. Fiz. 40, (1979) 15.
\bibitem{vil}R. Basu, A.H. Guth and A. Vilenkin, Phys. Rev. D44,
(1991) 340.
\bibitem{kri}I. Krichever, "Two-Dimensional Algebraic-Geometrical
Operators
with Self-Consistent Potential", to appear in Funct. Analis and Appl.
\bibitem{dev2}H.J. de Vega, A.L. Larsen and N. S\'{a}nchez, Phys.
Rev. D51,
             (1995) 6917.
\bibitem{rin}W. Rindler, Essential Relativity (Van Nostrand
              Reinhold CO, 1969).
\bibitem{abr}M. Abramowitz and I. Stegun, Handbook of mathematical
functions,
             ninth ed. (Dover, New York, 1972).
\bibitem{gsw}M.B. Green, J.H. Schwarz and E. Witten, Superstring
Theory, Vol 1
             (Cambridge University Press, Cambridge, 1986).
\bibitem{zel}V.P. Frolov, V. Skarzhinski, A. Zelnikov and O.
Heinrich,
             Phys. Lett. B224, (1989) 255.
\bibitem{vil2}A. Vilenkin, Phys. Rep. 121, (1985) 263.
\bibitem{fro}V.P. Frolov and N. S\'{a}nchez, Nucl. Phys. B349, (1991)
815.
\bibitem{car}S. Carlip, Class. Quantum Grav. 12, (1995) 2853.
\bibitem{ban2}M. Banados, M. Henneaux, C. Teitelboim and J. Zanelli,
Phys. Rev.
              D48, (1993) 1506.
\bibitem{wel}G.T. Horowitz and D.L. Welch, Phys. Rev. Lett. 71,
(1993) 328.
\end{thebibliography}
\end{document}